\documentclass[journal,draftcls,onecolumn,12pt,twoside, narroweqnarray]{IEEEtran}
\usepackage{color}
\usepackage{algorithm}
\usepackage{algorithmic}
\usepackage{graphicx}
\usepackage{amsmath,amssymb,amsthm}
\usepackage{latexsym}
\usepackage{cite}
\newtheorem{prop}{Proposition}
\newtheorem{Def}{Definition}
\newtheorem{Lemma}{Lemma}

\ifCLASSINFOpdf
   \graphicspath{{}{../pdf/}{../jpeg/}{../eps/}}
   \DeclareGraphicsExtensions{.jpg,.pdf,.jpeg,.png,.eps}
\else
\fi
\ifCLASSOPTIONcompsoc
\usepackage[caption=false,font=normalsize,labelfont=sf,textfont=sf]{subfig}
\else
\usepackage[caption=false,font=footnotesize]{subfig}
\fi
\usepackage{setspace}

\begin{document}
%
\title{Sensor Deployment with Limited Communication Range in Homogeneous and Heterogeneous Wireless Sensor Networks }
%
%
%
\author{Jun~Guo,~\IEEEmembership{Student Member,~IEEE,}
        and~Hamid~Jafarkhani,~\IEEEmembership{Fellow,~IEEE,}
\thanks{The authors are with Center for Pervasive Communications and Computing, University of California, Irvine
(e-mail: guoj4@uci.edu; hamidj@uci.edu).}}
\maketitle
\begin{abstract}
We study the heterogeneous wireless sensor networks (WSNs) and propose the necessary condition of the optimal sensor deployment. Similar to that in homogeneous WSNs, the necessary condition implies that every sensor node location should coincide with the centroid of its own optimal sensing region. Moreover, we discuss the dynamic sensor deployment in both homogeneous and heterogeneous WSNs with limited communication range for the sensor nodes. The purpose of sensor deployment is to improve sensing performance, reflected by distortion and coverage. We model the sensor deployment problem as a source coding problem with distortion reflecting sensing accuracy. Traditionally, coverage is the area covered by the sensor nodes. However, when the communication range is limited, a WSN may be divided into several disconnected sub-graphs. Under such a scenario, neither the conventional distortion nor the coverage represents the sensing performance as the collected data in disconnected sub-graphs cannot be communicated with the access point. By defining an appropriate distortion measure, we propose a Restrained Lloyd (RL) algorithm and a Deterministic Annealing (DA) algorithm to optimize sensor deployment in both homogeneous and heterogeneous WSNs. Our simulation results show that both DA and RL algorithms outperform the existing Lloyd algorithm when communication range is limited.
\end{abstract}


%
\IEEEpeerreviewmaketitle

\section{Introduction}
%
%
%
%
\IEEEPARstart{A}{s} a bridge between the physical world and the virtual information word, wireless sensor networks (WSNs) collect data from the physical world and communicate it with the virtual information world, such as computers. Proper sensor deployment improves monitoring and controlling the physical environment.
To accomplish their tasks, WSNs should address two needs: (i) Sensing in the target area and (ii) Communication between the sensor nodes.
WSNs are utilized to collect physical information, such as temperature, humidity, voice and so on. But, the collected data is useless if it cannot be transmitted to the access point (AP) and to the outside information world through the AP node. When sensors are connected by wire lines, the  connectivity is provided automatically.
On the other hand, the connectivity of WSNs is not guaranteed. In this paper, we consider the sensing and connectivity together and redefine the goal of WSN design accordingly.

A huge body of literature exists on the topic of sensor deployment. The sensor coverage range model assumes that sensors can only monitor the points within a range of $R_s$. The range $R_s$ is called the sensing range and the coverage area is the area covered by at least one sensor node \cite{SD}. Three different connectivity criteria are proposed in \cite{CC}. Three movement-assisted protocols, the VECtor-based algorithm, the VORonoi-based algorithm and the Minimax algorithm, are designed to maximize coverage area in \cite{wang1}. Lloyd algorithm has also been used as a tool to deploy sensors in homogeneous WSNs \cite{SD}. The convergence of the Lloyd algorithm has been studied in \cite{kiefferconvergence,wuconvergence,qdu1}. The analysis in [4]-[6] can be applied to the sensor deployment methods in [1].
However, \cite{SD} assumes an infinite communication range and ignores the connectivity limitation. When an infinite communication range is assumed, all the nodes in the network are connected to each other. In reality, each node has a limited communication range that will affect the connectivity of the network \cite{hu1,yousefizadeh1}. A geometric analysis of the relationship between the sensing coverage and the connectivity is proposed in \cite{ICC}. In \cite{DCC}, the authors have come up with some deployment patterns to achieve both sensing coverage and full connectivity.

Unfortunately, given a fixed number of sensor nodes and a finite communication range, connectivity is not guaranteed. When sensor nodes are divided into several disconnected sub-graphs, the conventional Lloyd algorithm cannot converge to a proper deployment. The Critical Sensor Density (CSD) in \cite{CSurvey} is the number of nodes per unit area, required to provide full sensing coverage when the communication range is limited.
When the sensor density is smaller than CSD, we cannot achieve the full sensing coverage. Under such a scenario, coverage may not be the right cost function to optimize. One needs to define an appropriate distortion measure to reflect the sensing accuracy. Distortion, as an important parameter in source coding can also be used to evaluate the WSN performance. Therefore, one can minimize distortion in WSNs through vector quantization techniques in \cite{gray1} and \cite{gershograybook}. The best possible distortion for a given number of sensors, i.e., the minimum distortion for a given rate, can be analyzed through the rate-distortion theory \cite{information theory}. Even if the sensor density is larger than CSD, there is no existing sensor deployment algorithm designed to achieve the full connectivity and minimize the distortion at the same time. In this paper, when the communication range is limited, we take both distortion and connectivity into consideration. The existing coverage area model is a special case of our distortion measure. We propose a method, named Restrained Lloyd (RL) Algorithm, to distribute sensor nodes and to minimize distortion with full connectivity. Then, a more complex approach, named Deterministic Annealing (DA) Algorithm, is designed to avoid sub-optimal solutions.

In many practical situations, different sensors in the WSN have different characteristics such as computational power, sensing range, and sensing accuracy.
The deployment and topology control of such heterogeneous WSNs that include the sensor nodes with different communication or sensing ranges, have been studied in \cite{mahboubi1} and \cite{HWS}. Similar to \cite{wang1}, three movement-assisted protocols in \cite{mahboubi1} are designed to avoid coverage hole in heterogeneous WSNs.
However, \cite{HWS} deploys sensor nodes one-by-one and to deploy a new sensor uses the location information of all previously deployed nodes. Also, \cite{HWS} assumes that sensor can monitor events within a circle with the radius equal to the sensing range. We generalize this model to a sensing accuracy which depends on the distance between the sensor and the event. Our distortion model will include the sensing range model in \cite{HWS} as a special case in which the distortion is a step function.
In such heterogeneous WSNs, weighted Voronoi diagrams \cite{VD} rather than conventional Voronoi diagrams \cite{voronoisurvey} will provide the best regions as we will discuss in this paper. An algorithm to construct weighted Voronoi diagrams for a different application has been suggested in \cite{aurenhammerwvd}. Based on the geometry of the optimal cell partitioning, we will analyze the objective functions in our model and propose the necessary condition for the optimal sensor deployment in heterogeneous WSNs with different sensing abilities.

In the rest of this paper, we first introduce the system model for both homogeneous and heterogeneous WSNs and formulate the problems of sensing and connectivity in Section \ref{sec:model}. Section \uppercase\expandafter{\romannumeral3} analyzes the optimal deployment in heterogeneous sensor networks without communication constraint. Section \uppercase\expandafter{\romannumeral4} proposes RL and DA algorithms to improve distortion and maintain connectivity. Section \uppercase\expandafter{\romannumeral5} presents simulation results and
Section \uppercase\expandafter{\romannumeral6} provides the conclusions.



\section{System Model and General Problems}\label{sec:model}
Let $Q$ be a simple convex polygon in $\Re^{2}$ including its interior. Given $n$ sensors in the target area $Q$, sensor deployment is defined by $P=(p_1,\cdots,p_n) \subset Q^{n}$, where $p_i$ is Sensor $i$'s location. For any point $q \in Q$, $\lambda(q)$ is the probability density function of an event at point $q$. A cell partition $R$ of $Q$ is a collection of disjoint subsets of $\{R_i(P)\}_{i\in{1,\cdots,n}}$ whose union is $Q$. Let $B(c,r)=\{q|\ \|q-c\|\le r\}$ be a disk centered at $c$ with radius $r$ in two-dimensional space. For two points $a$ and $b$, let equation $Eq+F=0$, where $E\in \Re^{2\times2}$ is a $2\times2$ matrix and $F\in \Re$ is a constant, define the perpendicular bisector hyperplane between the two points. Then, the equations $Eq+F\ge0$ and $Eq+F\leq 0$ define two half spaces. we denote the half space that contains point $a$ by $HS(a,b)$.

As mentioned before, we define the AP as the sensor node that can communicate with the outside information world.
Let $S(P)$ be the set of sensor nodes that can communicate with the AP when the sensor deployment is $P$. Note that in general not all nodes can communicate with the AP and  $card(S(P))\leq n$, where $card(A)$ is the number of elements in set $A$. We define a new sensor deployment, which is a subset of the all sensor locations, $H(P)$ as the vector of sensor locations for the $card(S(P))$ sensor nodes connected to the AP. When $S(P)$ includes all sensor nodes, we have $P=H(P)$ and $card(S(P))= n$. Let $T$ be the set of sensor deployments that provide full connectivity, i.e., $T=\{P|card(S(P))=n\}$. In our model, two sensor nodes can communicate with each other within one hop if and only if the distance between the two is smaller than $R_c$, where $R_c$ is referred to as the communication range.
A sensor node can transfer data outside if and only if there exists a path from the sensor to the AP. The path consists of a sequence of sensor nodes where each hop distance is smaller than the communication range $R_c$.
Sensor nodes that are connected to the AP construct the backbone network. If all sensors are included in the backbone network, we call the network fully connected. Otherwise, the network is divided into several disconnected sub-graphs.

Another important factor in analyzing the performance of a WSN is its sensing accuracy. Ideally, we would like to sense all events in the covered area. However, the sensing accuracy of a sensor node usually depends on the distance between the sensor and the event to be sensed.
In other words, the accuracy of the gathered data from an event at point $q$ by its associated sensor node $i$ is a non-increasing function of the distance between $q$ and $p_i$.
Therefore, to represent the average sensing accuracy in the target area, we define the following  general distortion:
\begin{equation}
    D(P) = \sum_{i=1}^n \int_{R_i(P)} \Phi_i(\|q-p_i\|) \lambda(q)dq,\label{D_P}
\end{equation}
where $\Phi_i(\cdot)$ is the cost function associated with sensing.
The partition $\{R_i(P)\}_{i\in1,\cdots,n}$ in the above definition include all sensor nodes. However, as explained before, when the communication range is limited, some sensor nodes cannot transfer their data back. As a result, only the sensor nodes in the backbone network can contribute to the sensing and therefore the distortion should be revised as
\begin{equation}
    D(P) = \sum_{i\in S(P)} \int_{R_i(H(P))} \Phi_i(\|q-p_i\|) \lambda(q)dq.\label{D_S_P}
\end{equation}
Note that to derive Eq. (\ref{D_S_P}) from Eq. (\ref{D_P}), one has to replace $P$ with $H(P)$, i.e., one has to consider only the sensor nodes that are in the backbone network.
We reiterate that in the case of a fully connected network, $H(P)=P$ and Eq. (\ref{D_P}) and Eq. (\ref{D_S_P}) are identical.

Obviously, choosing different cost functions in Eq. (\ref{D_S_P}) results in different problem formulations. One natural choice for the cost function is a continuous function defined by
\begin{equation}
    \Phi_i(x)=\eta_ix^2,
\end{equation}
where the cost parameter $\eta_i\in R^+$ is a constant that depends on the sensor characteristics. In homogeneous WSNs, every sensor node has the same sensing ability and the same cost parameter $\eta_i$. Therefore, the cost parameter can be ignored in homogeneous WSNs. However, different sensors with different complexity, power and sensing ability are used in heterogeneous WSNs. Obviously, the cost parameters $\{\eta_i\}_{i\in1,\cdots,n}$ reflect the quality of sensor nodes. The smaller the cost parameter, the stronger the sensing ability.

The distortion definitions in Eqs. (\ref{D_P}) and (\ref{D_S_P}) can represent the sensor coverage area model \cite{SD} as well. In such a model, the sensors can monitor events within a circle with a fixed radius called the sensing range. Consider a step function defined by
\begin{equation}
    I_i(x)=
    \begin{cases}
    1, &\mbox{for $\eta_ix^2<R_s^2$}\\
    0, &\mbox{for $\eta_ix^2\geq R_s^2$}.
    \end{cases}\label{step_cost}
\end{equation}
Adopting the above sensor coverage area model, choosing $\Phi_i(x)=I_i(x)$ in Eq. (\ref{D_S_P}) converts the distortion to the area covered by the sensors. In such a model, sensors can monitor events within a circle with the radius $\frac{R_s}{\sqrt{\eta_i}}$. Obviously, the coverage area should be maximized while the distortion should be minimized. Since by definition $\Phi_i(x)$ should be a non-decreasing function, to have the sensor coverage area model as a special case of our model, we should choose $\Phi_i(x)=1-I_i(x)$.

Our main goal is to minimize the distortion defined in Eq. (\ref{D_S_P}). 
It is easy to show that a necessary condition for such an optimal sensor deployment is to have a fully connected network. Moreover, the distortion is determined by both sensor deployment and cell partitioning. This is the topic of the  discussion in the next section.

\section{Optimal Deployment in Heterogeneous WSNs without Communication Constraint}
In this section, we assume an infinite communication range which results in a connected network for any sensor deployment. Given a fully connected network, the distortion $D(P)=D(P,R(P))$ is determined by the sensor deployment $P$ and the cell partition $R(P)$. In homogeneous WSNs, given sensors' locations, Voronoi partitions provide the smallest distortion. The Voronoi region (partition) for Sensor $i$, denoted by $V_i(P)$, is the intersection of half spaces $HS(p_i,p_j),\forall j\ne i$. In other words, the Voronoi region for Sensor $i$ is the set of all points that are closer to Sensor $i$ than any other sensor. The Voronoi partition of $Q$ generated by $P$ with respect to the Euclidean norm is the collection of sets $\{V_i(P)\}_{i\in{1,\cdots,n}}$ defined by
\begin{equation}
V_i(P)=\{q \in Q |\ \|q-p_i\| \le \|q-p_j\|, \forall j \in 1,\cdots,n\},
\end{equation}
where $\|\cdot\|$ is the Euclidean norm. This is because all sensors in homogeneous WSNs have the same cost function $\Phi(x)$ which is only determined by the Euclidean distance $x=\|q-p_i\|$. Since an event in $V_i(P)$ is monitored by Sensor $i$, each event is sensed by the nearest sensor and therefore makes the smallest contribution to the global distortion.

However, in a heterogeneous WSNs, the cost function $\Phi_i(x)$ is also affected by the cost parameter $\eta_i$, which reflects the sensing ability of Sensor $i$. Sensor nodes in heterogeneous WSNs can be classified according to their cost parameters $\{\eta_i\}_{i\in1,\cdots,n}$. Let us assume there are $m$ different sensor types with $m$ different cost parameters. Given sensor nodes' locations, an event at point $q$ should be sensed by the sensor with the smallest cost such that its contribution to the global distortion is the lowest possible. Such an optimal partitioning  is refer to as the weighted Voronoi partitioning. The weighted Voronoi partition of $Q$ generated by $P$ is the collection of sets ${V^H_i(P)}_{i\in{1,\cdots,n}}$ defined by
\begin{equation}
\label{df weighted Voronoi partition}
V^H_i(P)=\{q \in Q |\ \eta_i\|q-p_i\|^2 \le \eta_j\|q-p_j\|^2, \forall j \in 1,\cdots,n\}.
\end{equation}
Note that both Voronoi regions $V_i(P)_{i\in1,\cdots,n}$ and weighted Voronoi regions $V^H_i(P)_{i\in1,\cdots,n}$ are functions of $P$. Since the Voronoi partitioning can be considered as a special case of the weighted Voronoi partitioning, in which $\eta_i=1,i=1,\cdots,n$, we simply use $V^H_i(P)_{i\in1,\cdots,n}$ to represent both. Using the result of weighted Voronoi partitioning as the partition in Eq. (\ref{D_P}), the global distortion can be rewritten as
\begin{equation}
D(P) = \sum_{i=1}^n \int_{V^H_i(P)} \eta_i\|q-p_i\|^2 \lambda(q)dq.
\end{equation}
Our goal is to find the sensor deployment that minimizes the global distortion. In what follows, we generate the machinery to find the necessary condition of the optimal sensor deployment.

\begin{prop}
\label{Proposition.1}
Consider a two-dimensional heterogeneous sensor network, in which the cost function between Sensor $k$ and point $q$ is $\eta_k\|q-p_k\|^2$, the optimal partition for a given deployment $P$ is
\begin{equation}
\begin{aligned}
V^H_k(P) {=}&\:\{q \in Q |\ \eta_k\|q-p_k\| \le \eta_t\|q-p_t\|^2, \forall t \in 1,\cdots,n\}\\
         {=}&\:\left[\bigcap_{i:\eta_i<\eta_k}{B(c_{ik},r_{ik})}\right]\bigcap\left[\bigcap_{l:\eta_l=\eta_k}{HS(p_k,p_l)}\right] - \left[\bigcup_{j:\eta_j>\eta_k}{B(c_{kj},r_{kj})}\right],\\
\end{aligned}
\end{equation}
where $c_{ij}=\frac{p_j-(\eta_i/\eta_j)p_i}{1-(\eta_i/\eta_j)}$ and $r_{ij}=\frac{\sqrt{\eta_i/\eta_j}}{|1-\eta_i/\eta_j|}\|p_i-p_j\|$.
\end{prop}
\begin{IEEEproof}
The proof is provided in Appendix A.
\end{IEEEproof}
Before we discuss the optimal sensor deployment in heterogeneous WSNs, we need to present the following definitions and lemmas.
\begin{Def}
A set $S\subseteq \Re^n$ is called star-shaped if and only if there exists a point $p\in int(S)$ such that for all $s\in \partial S$ and all $\lambda \in (0,1]$, one has $\lambda p+(1-\lambda)s\in int(S)$, where $int(S)$ is the interior of $S$ and $\partial S$ is the boundary of $S$. The point $p$ is the reference point.
\end{Def}
\begin{Def}
A set $S\subseteq \Re^n$ is called a convex region if and only if for every pair of points $x,y\in S$ and all $\lambda \in (0,1)$, one has $\lambda x+(1-\lambda)y\in int(S)$.
\end{Def}
\begin{Lemma}
\label{a}
If a set $S\subseteq \Re^n$ is convex, then $S$ is star-shaped.
\end{Lemma}
\begin{IEEEproof}
For any convex region $S\subseteq \Re^n$, pick a point $p\in int(S)\subset S$. For any point $s\in \partial S\subset S$ and all $\lambda \in (0,1)$, one has $\lambda p+(1-\lambda)s\in S$. When $\lambda=1$, $\lambda p+(1-\lambda)s = p\in int(S)$. Therefore, the set $S$ is star-shaped.
\end{IEEEproof}
We will use the fact that the intersection of any collection of convex sets is convex \cite{Convexity,ACA} and as a result star-shaped according to Lemma \ref{a}.
\begin{Lemma}
\label{b}
The union of star-shaped sets that are associated with the same reference point $p$ is star-shaped.
\end{Lemma}
\begin{IEEEproof}
Given $m$ star-shaped sets $S_i, i=1,\cdots,m$ with the same reference point $p$, the corresponding union is $S=\bigcup_{i=1}^{m}S_i$. Since point $p$ is the reference point of star-shaped sets $S_i$, where $i\in\{1,\cdots,m\}$, we have $p\in\bigcap_{i=1}^{m}S_i$ and therefore $\bigcap_{i=1}^{m}S_i\ne\o$. The boundary of $S$ comes from the boundaries of $m$ star-shaped sets $S_i$, where $i\in\{1,\cdots,m\}$. Thus, for any point $s\in\partial S$, we have $s\in\bigcup_{i=1}^{m}\partial S_i$. Because of $m$ star-shaped sets, for all $s\in \partial S_i$ and all $\lambda\in(0,1]$, we will have $\lambda_ip+(1-\lambda)s\in int(S_i)$. Thus, for all $s\in \partial S$ and for all $\lambda\in(0,1]$, one can find a subset $S_i$ such that $s\in \partial S_i$ and so $\lambda p+(1-\lambda)s\in int(S_i)\subset int(S)$.
\end{IEEEproof}
\begin{Lemma}
\label{Lamma2}
Let $S(x)=\bigcup_{i=1}^mS_i(x)$ be a star-shaped set that consists of $m$ disjoint sub-sets $S_{i}$, where $i\in\{1,\cdots,m\}$. We then have
\begin{equation}
\int_{\partial S}\varphi(\gamma)n^t(\gamma)d\gamma=\sum_{i=1}^{m}\int_{\partial S_i}\varphi(\gamma)n^t(\gamma)d\gamma,
\end{equation}
where $\varphi(\cdot)$ is a continuous function of $\gamma$, and $n^t(q)$ is the unit outward normal to $\bigcup_{i=1}^m\partial S_i$ at $q$.
\end{Lemma}
\begin{IEEEproof}
\begin{equation}
\begin{aligned}
\sum_{i=1}^{m}\int_{\partial S_i}\varphi(\gamma)n^t(\gamma)d\gamma = \sum_{i=1}^{m}\left[\sum_{j\ne i}{\int_{\partial (S_i(x)\bigcap S_j)}\varphi(\gamma)n^t(\gamma)d\gamma} + \int_{\partial (S_i\bigcap S)}\varphi(\gamma)n^t(\gamma)d\gamma\right]
\end{aligned}
\end{equation}
For any $i$ and $j$ such that $S_i\bigcap S_j=\varnothing$, the corresponding curve integral $\int_{\partial (S_i\bigcap S_j)}\varphi(\gamma)n^t(\gamma)d\gamma$ is 0. On the other hand, for any $i$ and $j$ such that $S_i\bigcap S_j\ne\varnothing$, the corresponding curve integral $\int_{\partial (S_i\bigcap S_j)}\varphi(\gamma)n^t(\gamma)d\gamma=-\int_{\partial (S_j\bigcap S_i)}\varphi(\gamma)n^t(\gamma)d\gamma$ because of opposite unit outward normal. Therefore, we have
$\sum_{i=1}^{m}\int_{\partial S_i}\varphi(\gamma)n^t(\gamma)d\gamma
=\sum_{i=1}^{m}\left[\int_{\partial (S_i\bigcap S)}\varphi(\gamma)n^t(\gamma)d\gamma\right]
=\int_{\partial S}\varphi(\gamma)n^t(\gamma)d\gamma$
\end{IEEEproof}
\begin{Lemma}
\label{Lamma3}
Let $S=\bigcup_{i=1}^mS_i$ be a star-shaped set that consists of $m$ disjoint subsets $S_i$, $i=1,\cdots,m$. Then for any point $p\in S$ we have
\begin{equation}
(p-C_S)M_s=\sum_{i=1}^{m}(p-C_{S_i})M_{s_i},\label{pCM}
\end{equation}
where $M_S=\int_S\lambda(q)dq$ and $C_S=\frac{\int_Sq\lambda(q)dq}{M_S}$ are, respectively, the mass and the center of of mass with respect to the probability density function $\lambda(\cdot)$ of the set S.
\end{Lemma}
\begin{IEEEproof}
We rewrite the left side of Eq. (\ref{pCM}) to derive
\begin{equation}
\begin{aligned}
(p-C_S)M_s
&{=}pM_S-\sum_{i=1}^{m}\int_{S_i}q\lambda(q)dq\\
&{=}\sum_{i=1}^{m}\left[pM_{S_i}-\int_{S_i}q\lambda(q)dq\right]\\
&{=}\sum_{i=1}^{m}(p-C_{S_i})M_{s_i}\\
\end{aligned}
\end{equation}
\end{IEEEproof}
Now, we have enough tools to derive the main results in this section.
Proposition A.1. in \cite{SD} presents how to calculate the gradient of the distortion when sensing regions are star-shaped. Unfortunately, it is possible that the sensing regions in heterogeneous WSNs are not star-shaped. In what follows, we show how to calculate the gradient of the distortion in heterogeneous WSNs.
\begin{prop}
\label{prop2}
In a heterogeneous sensor network including $m$ kinds of sensors, let $P=\left[p_1,p_2,\cdots,p_n\right]$ be the sensor deployment, and $W\in\Re^2$ be an arbitrary convex set. Let a series of functions $\varphi_i:\Re^2\times(a,b)\to \Re$, where $i=1,\cdots,n$, be continuous on $\Re^2\times(a,b)$, continuously differentiable with respect to its second argument for all $p_i\in(a,b)^2$, where $i\in{1,\cdots,n}$, and almost all $q\in\Re^2$, and such that for each $p_i\in(a,b)^2$, the maps $q\mapsto \varphi_i(q,p_i)$ and $q\mapsto \frac{\partial \varphi_i}{\partial x}(q,x)$ are measurable, and integrable on $\Re^2$. Then the function
\begin{equation}
\int_{V^H_i(P)\bigcap W}\varphi_i(q,p_i)dq
\end{equation}
is continuously differentiable and
\begin{equation}
\begin{aligned}
\label{EQ1}
&\:\frac{\partial \int_{V^H_i(P)\bigcap W}\varphi_i(q,p_i)dq}{\partial p_j} = \int_{V^H_i(P)\bigcap W}\frac{\partial \varphi_i(q,p_i)}{\partial p_j}dq + \int_{\partial\left[V^H_i(P)\bigcap W\right]}\varphi_i(\gamma,p_i)n^t(\gamma)\frac{\partial\gamma}{\partial p_j}dq.
\end{aligned}
\end{equation}
\end{prop}
\begin{IEEEproof}
The proof is provided in Appendix B.
\end{IEEEproof}
Note that the sensing cell $V^H_i(P)$ in Proposition \ref{prop2} is a weighted Voronoi region and different from the Voronoi region in \cite{SD}. The weighted Voronoi region $V^H_i(P)$ can be a non-star-shaped region and therefore we need to use Proposition \ref{prop2}.

Next, we derive the necessary condition for the optimal deployment in heterogeneous WSNs when the communication range is infinite. The format of the result, as proved in the next proposition, is similar to that of homogeneous WSNs.
\begin{prop}
When the communication range is infinite, the necessary condition for the optimal deployment in heterogeneous WSNs is
\begin{equation}
p_j^* = c_j(P), \forall j\in\{1,\cdots,n\},
\label{critical deployment}
\end{equation}
where $p_j^*$ is the optimal position for sensor node $j$, $M_j(P)=\int_{V^H_j(P)}\lambda(q)dq$ and $c_j(P)=\frac{\int_{V^H_j(P)}q\lambda(q)dq}{M_j(P)}$ are,respectively, the mass and the center of weighted Voronoi cell $V^H_j(P)$ with respect to the probability density function $\lambda(\cdot)$ in target region $Q$.
\end{prop}
\begin{IEEEproof}
Let $W=\Re^2$ and $\varphi_i(q,p_i)=\Phi_i(\|q-p_i\|)\lambda(q)$, where $\Phi_i(\cdot)$ is the cost function, we can use Proposition \ref{prop2} to calculate the partial derivative of the local distortion as follows:
\begin{equation}
\begin{aligned}
&\frac{\partial\int_{V^H_i(P)}\eta_i\|q-p_i\|^2\lambda(q)dq}{\partial p_j} =
\begin{cases}
2\eta_i(p_i-c_i)M_i + \int_{\partial V^H_i(P)}\eta_i\|\gamma-p_i\|^2n^t(\gamma)\frac{\partial \gamma}{\partial p_i}d\gamma,&i=j\\
\int_{\partial V^H_i(P)}\eta_i\|\gamma-p_i\|^2n^t(\gamma)\frac{\partial \gamma}{\partial p_j}\lambda(\gamma)d\gamma,&i\ne j
\end{cases}
\end{aligned}\label{partial}
\end{equation}
where $\frac{\partial \gamma}{\partial p_j}\ne0$ if and only if $\gamma$ is on the boundary of $V^H_j(P)$.

Note that $\partial V^H_i(P)=\bigcup \partial \left[V^H_i(P)\bigcap V^H_l(P)\right], \forall l\ne i$.
  Thus, in Eq. (\ref{partial}) only the curve integral for $j$, i.e.,  on $\partial \left[V^H_i(P)\bigcap V^H_j(P)\right]$, needs to be taken into account. For each curve integral in the second case ($i\ne j$), one can find a curve integral with the opposite unit outward normal in the first case ($i=j$). As a result, the curve integrals in the partial derivative of the global distortion are canceled with each other. Therefore,
\begin{equation}
\begin{aligned}
\frac{D(P)}{\partial p_j} &= \sum_{i=1}^{n}\frac{\partial \int_{V^H_i(P)}\eta_i\|q-p_i\|^2\lambda(q)dq}{\partial p_j}\\
&= 2\eta_j(p_j-c_j(P))M_j(P), j=1,\cdots,n.
\end{aligned}
\end{equation}
Both $M_j(P)$ and $c_j(P)$ are functions of $P$. The optimal deployment $P^{\ast}$ will have a zero gradient. We define the cost parameters to be positive. Thus, when the communication range is infinite, the necessary condition for optimal deployment is the same as Eq. (\ref{critical deployment}).
\end{IEEEproof}
\section{Restraint Lloyd Algorithm and Deterministic Annealing Algorithm}
In this section, we design algorithms to minimize the distortion when the communication range is limited.
First, we quickly review the conventional Lloyd algorithm. Lloyd Algorithm has two basic steps in each iteration: (1) Sensor nodes move to their centroid; (2) Partitioning is done by assigning the corresponding Voronoi region to each sensor node. Lloyd Algorithm provides good performance and is simple enough to be implemented distributively. It converges to a minimum distortion when the communication range is infinite \cite{SD}. Unfortunately, it also has three shortcomings. First, since minimizing distortion is a non-convex optimization problem, Lloyd Algorithm may end at a large local minimum point rather than the optimal global minimum. Second, Lloyd Algorithm may not result in a connected network. Third, when WSNs are divided into several disconnected sub-graphs, Lloyd Algorithm is not feasible. In other words, since there is no global information available about the sensor locations, each sub-graph will run the algorithm independently. To deploy a network with full connectivity and lower distortion, we add some restraints on sensors' movements. We design a class of algorithms based on the Lloyd algorithm, referred to as RL Algorithm.

\subsection{Restrained Lloyd Algorithm}
Before we introduce the details of our RL Algorithm, we introduce the concept of a desired region. Let us assume we are trying to move Sensor $i$ at a given step.
Our goal is to keep the connectivity of the backbone network after moving Sensor $i$. Therefore, we define the areas in which Sensor $i$ will be connected to the backbone network as its desired region, denoted by $L_i(P)$. Note that this region may not be a star-shaped set. In our RL Algorithm, if Sensor $i$ is in the backbone network, we will restrain its movement within its desired region. To achieve this goal, we need to find the desired region $L_i(P)$.
Given a deployment $P$, if Sensor $i$ from the backbone network is removed, the rest of the sensor nodes in the backbone network will be divided into $K_i$ components: $U_{i1}(P),U_{i2}(P),\cdots,U_{iK_i}(P)$, where $U_{ij}(P)$ is a set of sensors included in the $j$th component. Note that $K_i$ may be equal to one. Then, we can calculate the desired  region as
\begin{equation}
    L_i(P) = \bigcap_{k=1}^{K_i}\left[\bigcup_{j\in U_{ik}}B(p_j,R_c)\right].
\end{equation}
Since the desired region is primarily influenced by the neighboring sensor nodes, we can approximate it by
\begin{equation}
    \tilde{L}_i(P) = \bigcap_{k=1}^{K_i}\left[\bigcup_{j\in U_{ik}\bigcap N_i(P)}B(p_j,R_c)\right],\label{LiP}
\end{equation}
where $N_i(P)$ consists of Sensor $i$'s neighbors when the deployment is $P$. Note that the approximation in (\ref{LiP}) can be calculated locally, but to calculate the exact desired region, one needs global information. Also, according to Lemma \ref{b}, the approximate desired region $\tilde{L}_i(P)$ is a star-shaped set.

Now, we provide the details of our RL Algorithm. The algorithm iterates between two steps: \\
(1) Sensors in the backbone network move one by one. Every sensor in the backbone network calculates its own approximate desired region $\tilde{L}_i(P)$ and moves to a location which is the closest point to its centroid $c_i(P)$ within $\tilde{L}_i(P)$. Sensors outside the backbone network move randomly and check if there is a path to the AP. Unlike the conventional Lloyd algorithm, these new locations may not be the centroid of the partition regions;
\\
(2) The target area, $Q$, is partitioned to weighted Voronoi regions for sensors in the backbone network, $S(P)$.

The main difference between RL Algorithm and the Lloyd algorithm is in the first step.
In what follows, we show that Step (1) in RL Algorithm will not increase the global distortion.


The global distortion is the sum of local distortions defined by
\begin{equation}
D_i(P) = \int_{V^H_i(H(P))}\|q-p_i\|^2\lambda(q)dq, i\in\{1,\cdots,m\},
\end{equation}
where $m$ is the number of sensors in the backbone network.
The local distortion, whether in homogeneous WSNs or heterogeneous WSNs, is a convex function. According to the parallel axis theorem, the local distortions can be rewritten as
\begin{equation}
\begin{aligned}
\label{aa}
D_i(P) =& \int_{V^H_i(H(P))}\|q-c_i(P)\|^2\lambda(q)dq + \int_{V^H_i(H(P))}\lambda(q)dq \cdot \|p_i-c_i(P)\|^2, i\in\{1,\cdots,m\},
\end{aligned}
\end{equation}
where $c_i(P) = \frac{1}{\int_{V^H_i(H(P))}\lambda(q)dq}\int_{V^H_i(H(P))}q\lambda(q)dq$ is the centroid of the partition region $V^H_i(H(P))$ with respect to the probability density function. Both $\int_{V^H_i(H(P))}\|q-c_i(P)\|^2\lambda(q)dq$ and $\int_{V^H_i(H(P))}\lambda(q)dq$ are constants when the integral area $V^H_i(H(P))$ is fixed. In other words, the local distortion is directly proportional to $\|p_i-c_i(P)\|^2$, i.e., the  sensor's distance to its centroid. Therefore, the movements in Step (1) minimize the local distortions. As the sum of local distortions, the global distortion will not increase. Since the sequence of the global distortion values is a non-increasing sequence with a lower bound of zero, it will converge.


We also show that our RL Algorithm guarantees the connectivity of the network with high probability after enough number of iterations.
Note that once a sensor node finds a path to the AP, our RL Algorithm will keep it in the backbone network.
Intuitively, as we have more iterations, the sensors outside the backbone network will move randomly and eventually connect to the AP as well.
Quantitatively, for the deployment after $k$ iterations, the area in which a sensor can communicate with the backbone network can be calculated by  $A_k=\mbox{AREA}\left(Q\bigcap\left[\bigcup_{i\in backbone}B(p_i,R_c)\right]\right)$. 
Then, the probability that a sensor outside the backbone network is not connected to the AP in its next move is $\frac{\mbox{AREA}(Q)-A_k}{\mbox{AREA}(Q)}$.
After $N$ iterations, the probability that a sensor is still out of the backbone network can be calculated by $P_{out}(N)=\prod_{k=1}^{N}\left[\frac{\mbox{AREA}(Q)-A_k}{\mbox{AREA}(Q)}\right]\leq\left[\frac{\mbox{AREA}(Q)-\min{A_k}}{\mbox{AREA}(Q)}\right]^N$ and then $\lim_{N\to\infty}P_{out}(N)=0$ because of $\min{A_k}>0$. In other words, as long as the number of iterations is large enough, almost all sensor nodes will be included in the backbone network, indicating full connectivity, with high probability.
\subsection{Deterministic Annealing Algorithm}
Like any other steepest-descent algorithm, RL Algorithm may converge to a local minimum a large distortion. One approach to improve the sub-optimal solution or find the global optimal solution, is to use annealing methods. Simulated Annealing (SA) \cite{SA,SR} is a method in which a candidate sensor movement is generated randomly. However, SA ignores the characteristics of the objective function and requires burdensome computations. In this paper, we design a DA algorithm which combines RA with annealing to minimize the distortion. Unlike SA, the proposed DA generates two new sensor positions deterministically at each iteration; however, choose one of the two options randomly.
Like RL Algorithm, our DA Algorithm iterates between two steps. The second step is identical to that of RL Algorithm. In the first step, the algorithm creates two candidate locations for each node in the backbone network. One candidate is the RL Algorithm's candidate that minimizes the local distortion. On the other hand, the second candidate increases the local distortion. It is easy to show that to maximize the local distortion for Sensor $i$ in the backbone network, one should move it to the point $o$ on the boundary of the desired region $\tilde{L}_i(P)$ that has the largest distance to the centroid $c_i(P)$. But the goal of the second candidate is to increase the distortion and not necessarily maximizing it. Moreover, the distortion is more sensitive to the sensors with smaller cost parameters. In order to avoid increasing distortion too fast, Sensor $i$ moves to the point $p_i+\frac{\eta_i}{\min_j(\eta_j)}(o-p_i)$. The algorithm will choose the first candidate with probability $p$ and increases $p$ from 0 to 1. Otherwise, the algorithm will choose the second candidate. In our algorithm, the probability $p$ is increased in proportion to $\log k$, where $k$ is the iteration count and for the last $M$ iterations we force the probability $p=1$.
Like RL Algorithm, DA Algorithm guarantees connectivity and convergence. The proof is similar to that of RL Algorithm and is omitted.

\section{Performance Evaluation}
We compare the performance of RL Algorithm, DA Algorithm and Lloyd Algorithm in sensor networks. We provide simulations in three sensor networks: (1) WSN1: A homogeneous WSN in which all sensors have the same cost parameter $\eta_i=1, i=1,\cdots,16$; (2) WSN2: A heterogeneous WSN including 2 kinds of sensors: four strong sensors with $\eta_i=1$ and twelve weak sensors with $\eta_j=16$; (3) WSN3: A heterogeneous WSN including three kinds of sensors: two strong sensors with $\eta_i=1$, four medium sensors with $\eta_j=4$ and ten weak sensors with $\eta_k=16$. Sixteen sensors are provided in each sensor network.  The AP is chosen from the sixteen sensors randomly. However, when we report the distortion or coverage area for the Lloyd algorithm, we report that of the largest connected subgraph which may not be connected to the AP. Obviously, this will be advantageous for the Lloyd algorithm, but our proposed algorithms still outperform the Lloyd algorithm. We use ten random initial deployments for each algorithm. To have a fair comparison, we consider the same target domain $Q$ as in \cite{SD}. $Q$ is determined by the polygon vertices
{(0,0), (2.125,0), (2.9325,1.5), (2.975,1.6), (2.9325,1.7), (2.295,2.1), (0.85,2.3), (0.17,1.2)}. 
The distribution of the events is also the same as \cite{SD}. The probability density function is the sum of five Gaussian functions of the form $5exp(6(-(x-x_{center})^2-(y-y_{center})^2))$. The centers $(x_{center},y_{center})$ are (2,0.25), (1,2.25), (1.9,1.9), (2.35,1.25) and (0.1,0.1). We use 0.5 as the communication range $R_c$. Also, when reporting the coverage area using (\ref{step_cost}), we use 0.25 as the sensing range $R_s$. In DA Algorithm, the first candidate is accepted at the $i$th iteration by a probability of $p(i)=log(i+1)/log(N+1)$, where $N$ is the number of regular iterations. Additional $M=25$ iterations are used in DA Algorithm to avoid ending with a process that increases the local distortions.
\begin{figure*}[!t]
\centering
\subfloat[]{\includegraphics[width=2.7in]{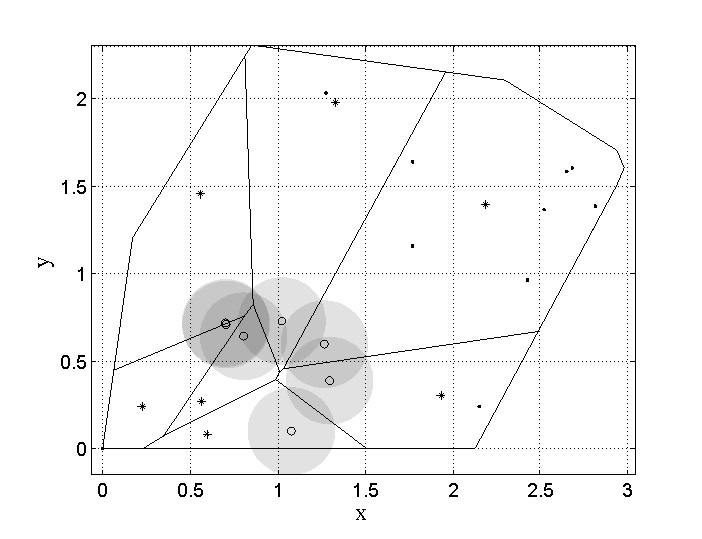}
\label{Lloyd initial deployment}}
\hfil
\subfloat[]{\includegraphics[width=2.7in]{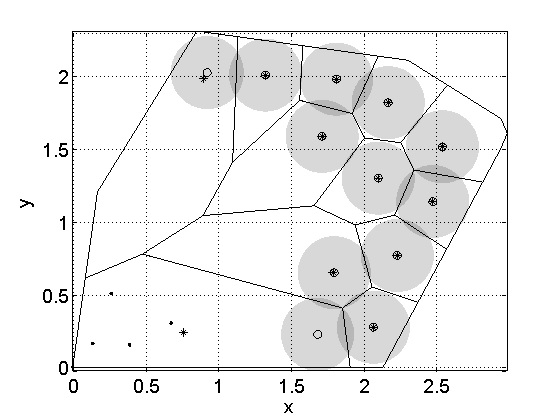}
\label{Lloyd final deployment}}
\hfil
\subfloat[]{\includegraphics[width=2.7in]{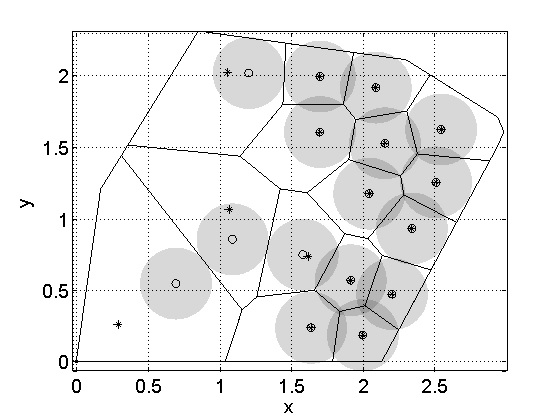}
\label{RL final deployment}}
\hfil
\subfloat[]{\includegraphics[width=2.7in]{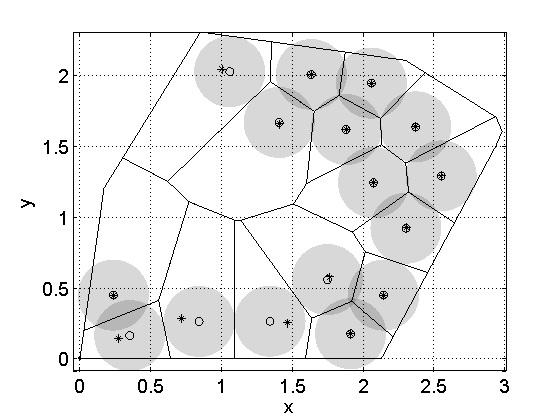}
\label{DA final deployment}}
\caption{Sensor deployments in WSN1. (a) The initial sensor deployment and the corresponding Voronoi regions. (b) The final deployment of Lloyd Algorithm after 500 iterations. (c) The final deployment of RL Algorithm after 500 iterations. (d) The final deployment of DA Algorithm after 500 iterations. Sensors in the backbone network are marked by circles. Sensors disconnected from the backbone network are denoted by dots. Voronoi region centroids are marked by stars. The radius of each gray circle is $R_c/2=0.25$.}
\label{Deployment In WSN1}
\end{figure*}

Figs. \ref{Lloyd initial deployment} and \ref{Lloyd final deployment} show one example of the initial and the finial deployments of Lloyd Algorithm in WSN1. Lloyd Algorithm assumes an  infinite communication range and requires the global knowledge of the sensor locations. Otherwise, disconnected sub-graphs run Lloyd Algorithm independently and there is no guarantee for convergence. Nonetheless, the calculation of the final distortion only considers sensors in the backbone network. In the final deployment of the example in Fig. \ref{Lloyd final deployment}, there are four sensors disconnected from the backbone network, resulting in a large distortion $D(P)=2.21$.
Fig. \ref{RL final deployment} shows the outcome of RL Algorithm in WSN1. After 500 iterations, the distortion is decreased from 11.30 to 0.60. Simultaneously, the coverage area is increased from 0.15 to 6.26 and the final deployment is connected.
Fig. \ref{DA final deployment} shows the final deployment of DA Algorithm in WSN1. After 500 iterations, the distortion is decreased from 11.30 to 0.32, which is better than that of RL Algorithm. Simultaneously, the coverage area is increased from 0.15 to 6.99 and full connectivity is provided. Unlike Lloyd Algorithm, both RL Algorithm and DA Algorithm guarantee connectivity.

\begin{figure}[!t]
\centering
\includegraphics[width=4.5in]{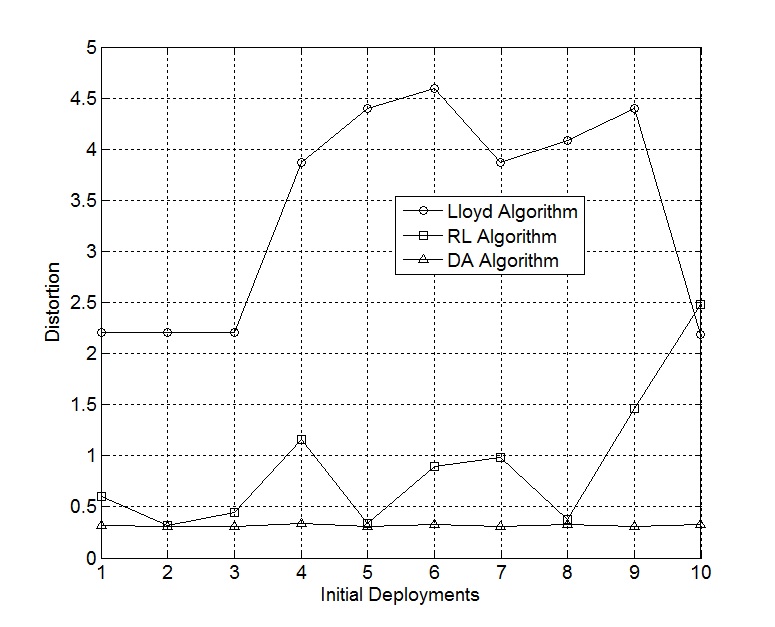}
\caption{Comparison of distortion for different algorithms in WSN1.}
\label{distortion in homogeneous sensor network}
\end{figure}
\begin{figure}[!t]
\centering
\includegraphics[width=4.5in]{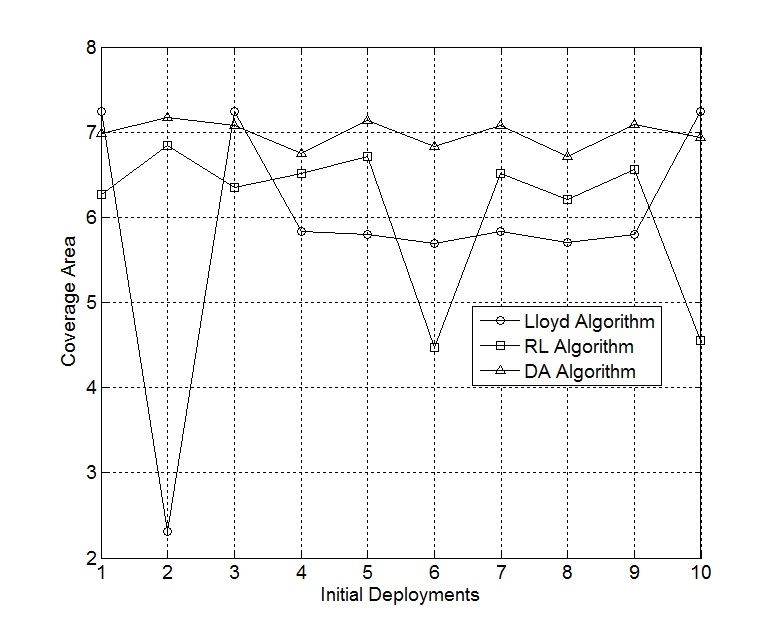}
\caption{Comparison of coverage for different algorithms in WSN1.}
\label{coverage in homogeneous sensor network}
\end{figure}

\begin{figure}[!t]
\centering
\includegraphics[width=4.5in]{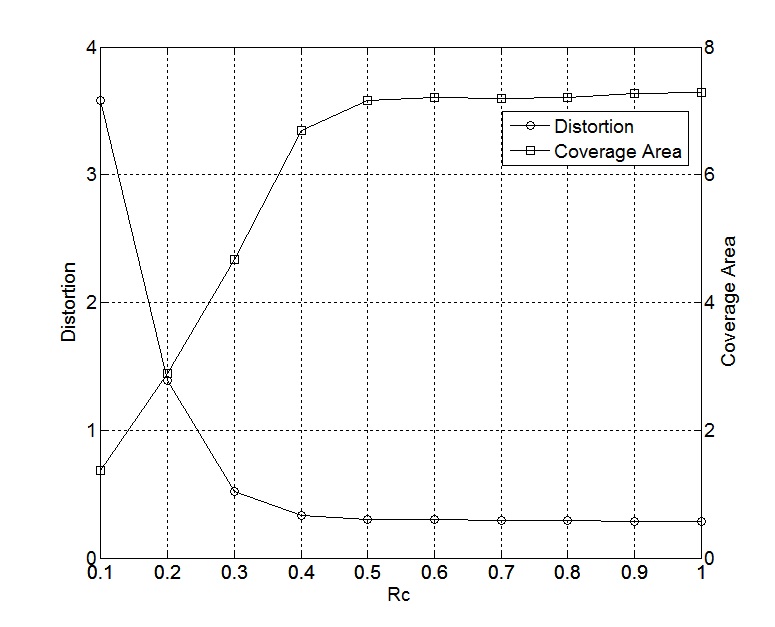}
\setlength{\belowcaptionskip}{-10pt} \caption{Relationship between performance and Rc in WSN1.}
\label{relationship between distortion, coverage and Rc}
\end{figure}

Fig. \ref{distortion in homogeneous sensor network} illustrates the performance of the above algorithms for 10 random initial deployments. As can be seen from the figure, unlike other algorithms, the performance of DA Algorithm is not sensitive to the initial deployment. In other words, DA Algorithm avoids most poor local minimum solutions. Fig. \ref{distortion in homogeneous sensor network} shows that DA Algorithm has the best performance among the three algorithms.
\begin{figure*}[!t]
\centering
\subfloat[]{\includegraphics[width=2.7in]{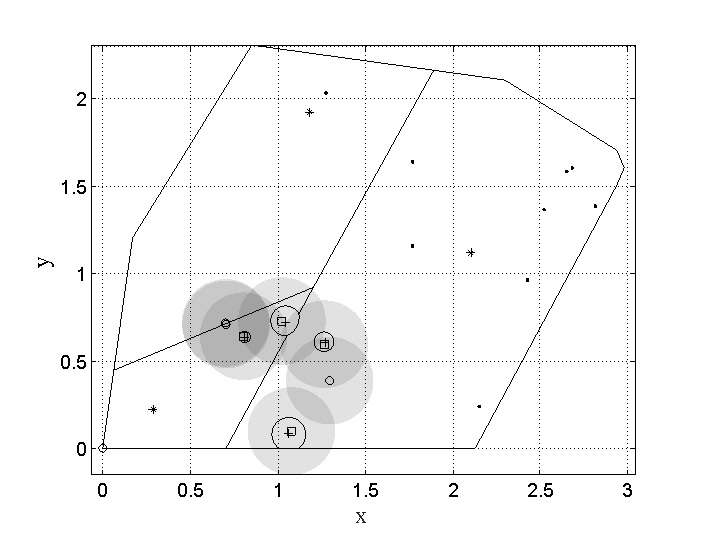}
\label{Lloyd initial deployment in WSN2}}
\hfil
\subfloat[]{\includegraphics[width=2.7in]{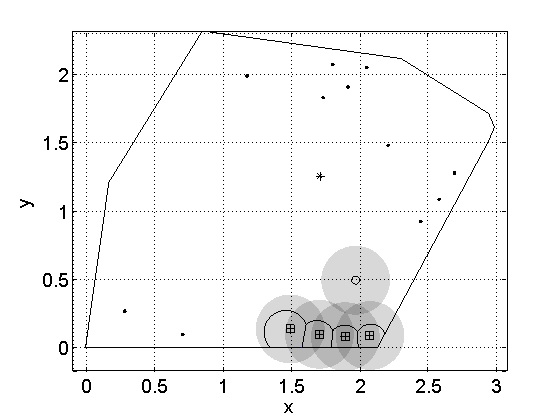}
\label{Lloyd final deployment in WSN2}}
\hfil
\subfloat[]{\includegraphics[width=2.7in]{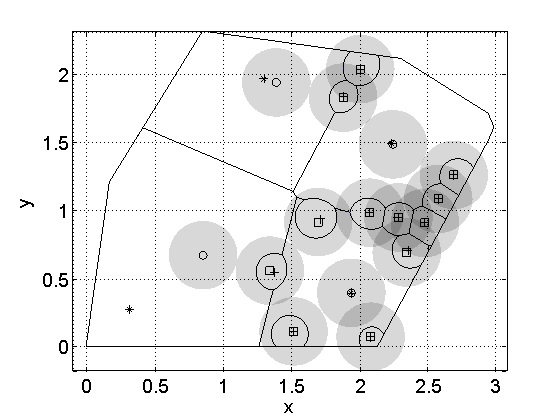}
\label{RL final deployment in WSN2}}
\hfil
\subfloat[]{\includegraphics[width=2.7in]{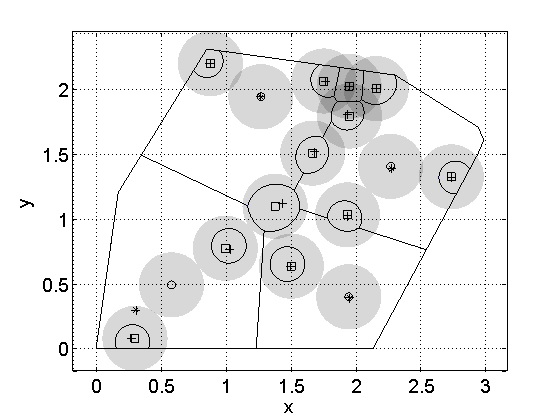}
\label{DA final deployment in WSN2}}
\caption{Sensor deployments in WSN2. (a) The initial sensor deployment and the corresponding weighted Voronoi regions. (b) The final deployment of Lloyd Algorithm after 500 iterations. (c) The final deployment of RL Algorithm after 500 iterations. (d) The final deployment of DA Algorithm after 500 iterations. Strong sensors and weak sensors in the backbone network are, respectively, denoted by hollow circles and squares. Sensors out of the backbone network are denoted by dots. The corresponding centroid for strong sensors and weak sensors are, respectively, denoted by stars and crosses. The radius of each gray circle is $R_c/2=0.25$.}
\label{DA Deployment in WSN2}
\end{figure*}
\begin{figure*}[!t]
\centering
\subfloat[]{\includegraphics[width=2.7in]{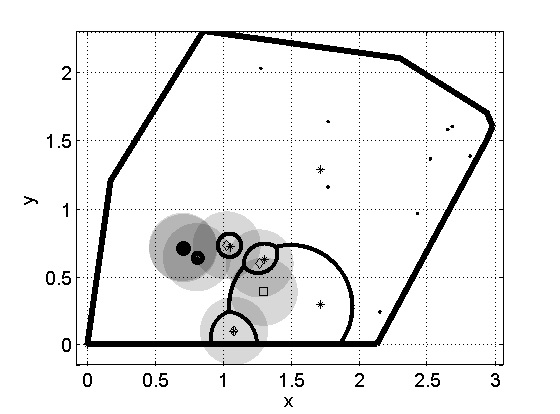}
\label{Lloyd initial deployment in WSN3}}
\hfil
\subfloat[]{\includegraphics[width=2.7in]{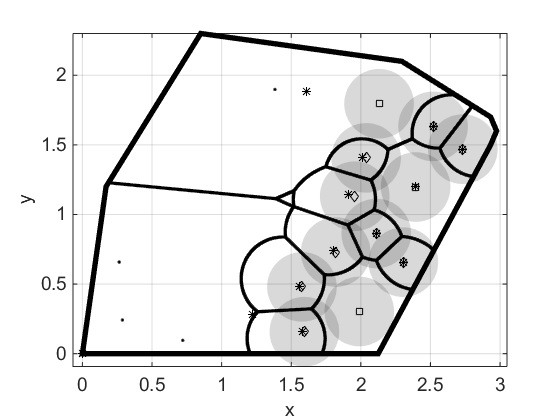}
\label{Lloyd final deployment in WSN3}}
\hfil
\subfloat[]{\includegraphics[width=2.7in]{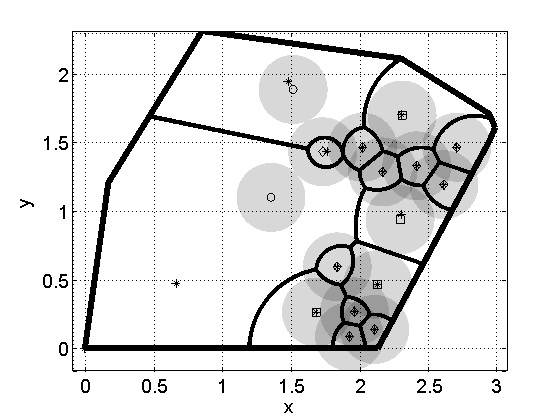}
\label{RL final deployment in WSN3}}
\hfil
\subfloat[]{\includegraphics[width=2.7in]{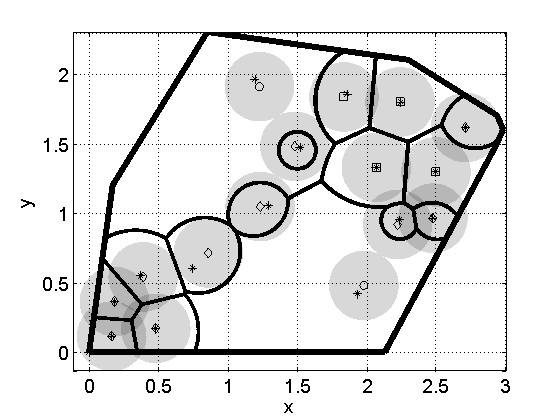}
\label{DA final deployment in WSN3}}
\caption{Sensor deployments in WSN3. Figure (a) The initial deployment and corresponding weighted Voronoi regions. (b) The final deployment of Lloyd Algorithm after 500 iterations. (c) The final deployment of RL Algorithm after 500 iterations. (d) The final deployment of DA Algorithm after 500 iterations. Strong sensors, medium and weak sensors in the backbone network are, respectively, denoted by hollow circles, hollow squares and hollow diamonds. Sensors out of the backbone network are denoted by dots. The corresponding centroid for sensors in the backbone network are denoted by stars. The radius of each gray circle is $R_c/2=0.25$.}
\label{DA Deployment in WSN3}
\end{figure*}

Fig. \ref{coverage in homogeneous sensor network} compares the final coverage area of RL Algorithm and DA Algorithm with that of Lloyd Algorithm. In most cases, decreasing the distortion results in increasing the coverage area as well. Intuitively, this behavior can be explained by considering coverage area as a hard-decision version of distortion. Next, the relationship between performance (distortion and coverage area) and communication range $R_c$ in homogeneous WSN1 using DA Algorithm is depicted in Fig. \ref{relationship between distortion, coverage and Rc}.

Figs. \ref{Lloyd initial deployment in WSN2} and \ref{Lloyd final deployment in WSN2} show one example of the initial and the finial deployments of Lloyd Algorithm in WSN2.
As usual, the final distortion only considers sensors in the backbone network.
Initially, two strong sensors and four weak sensors are consisted in the backbone network shown in Fig. \ref{Lloyd initial deployment in WSN2}. In Fig. \ref{Lloyd final deployment in WSN2}, only one strong sensor and four weak sensors are included in the backbone network, resulting in a large distortion $D(P)=12.67$ which is only 0.48 smaller than the initial distortion.
Fig. \ref{RL final deployment in WSN2} shows the outcome of RL Algorithm in WSN2. After 500 iterations, the distortion is decreased from 13.15 to 5.52. Simultaneously, the coverage area is increased from 0.08 to 1.46 and the final deployment is connected.
Fig. \ref{DA final deployment in WSN2} shows the final deployment of DA Algorithm in WSN2. After 500 iterations, the distortion is decreased from 13.15 to 1.10, which is better than that of RL Algorithm. Simultaneously, the coverage area is increased from 0.08 to 2.48 and full connectivity is provided.

Figs. \ref{Lloyd initial deployment in WSN3} and \ref{Lloyd final deployment in WSN3} show one example of the initial and the finial deployments of Lloyd Algorithm in WSN3.
Initially, one strong sensor, two medium sensors and four weak sensors are consisted in the backbone network shown in Fig. \ref{Lloyd initial deployment in WSN3}. In Fig. \ref{Lloyd final deployment in WSN3}, two strong sensors, one medium sensors and one weak sensors are disconnected from the backbone network, resulting in a large distortion $D(P)=19.83$.
Fig. \ref{RL final deployment in WSN3} shows the outcome of RL Algorithm in WSN3. After 500 iterations, the distortion is decreased from 10.96 to 3.22. Simultaneously, the coverage area is increased from 0.08 to 1.51 and the final deployment is connected.
Fig. \ref{DA final deployment in WSN3} shows the final deployment of DA Algorithm in WSN3. After 500 iterations, the distortion is decreased from 10.96 to 1.35, which is better than that of RL Algorithm. Simultaneously, the coverage area is increased from 0.03 to 2.07 and full connectivity is provided.


\begin{figure}[!t]
\centering
\includegraphics[width=4.5in]{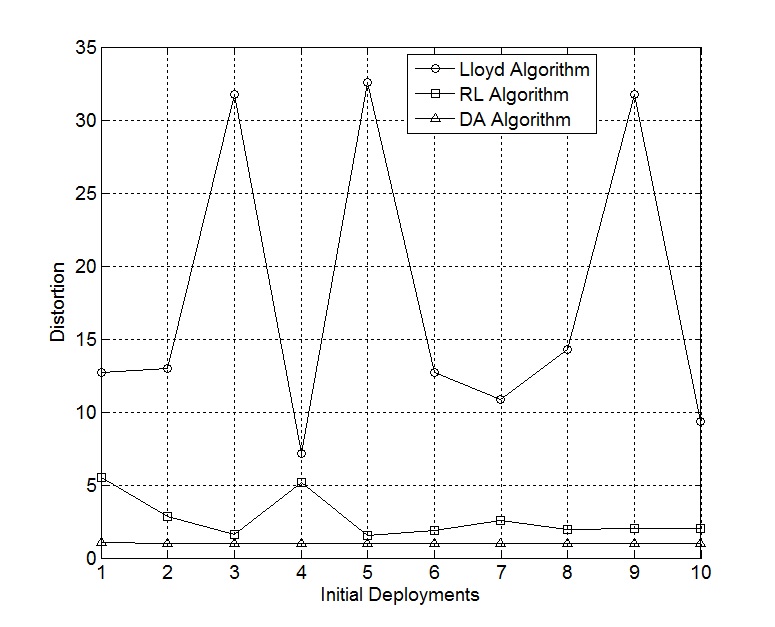}
\setlength{\belowcaptionskip}{-10pt} \caption{Comparison of distortion with different algorithms in WSN2.}
\label{distortion in heterogeneous sensor network2}
\end{figure}
\begin{figure}[!t]
\centering
\includegraphics[width=4.5in]{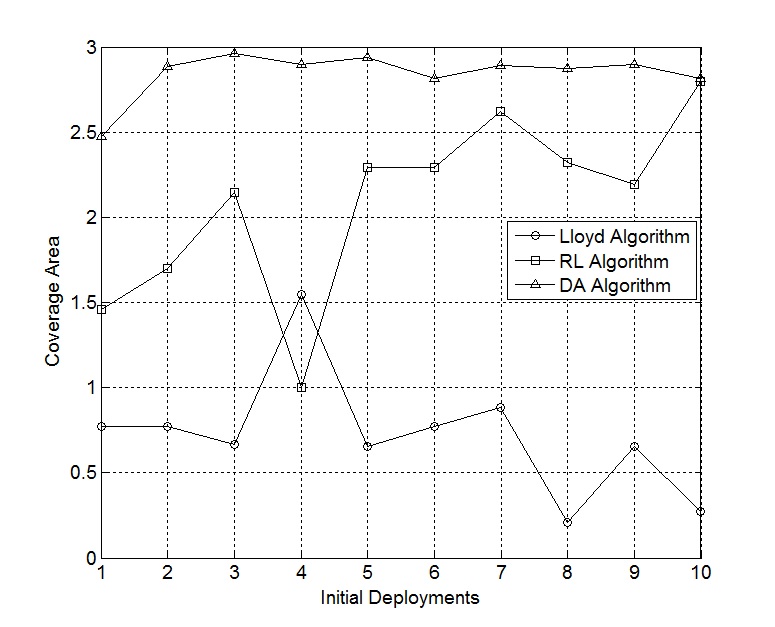}
\setlength{\belowcaptionskip}{-10pt} \caption{Comparison of coverage area with different algorithms in WSN2.}
\label{coverage in heterogeneous sensor network2}
\end{figure}

\begin{figure}[!t]
\centering
\includegraphics[width=4.5in]{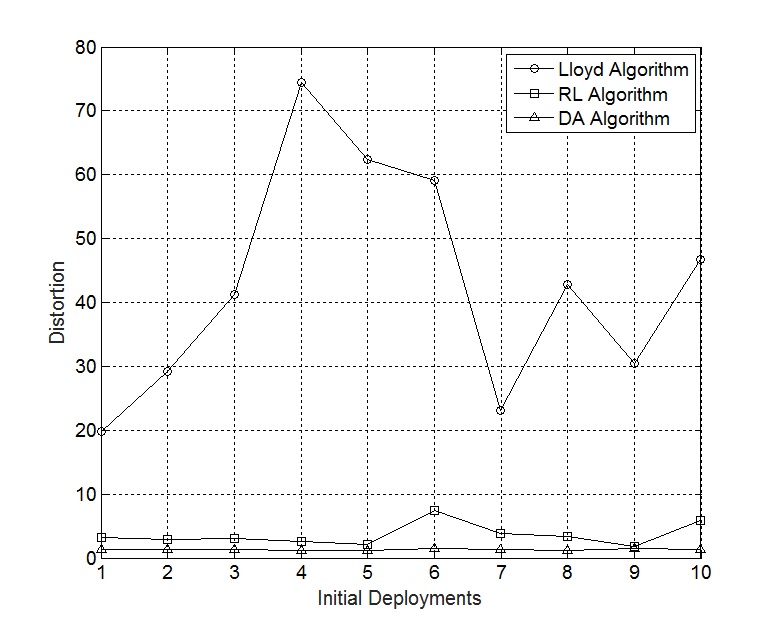}
\setlength{\belowcaptionskip}{-10pt} \caption{Comparison of distortion with different algorithms in WSN3.}
\label{distortion in heterogeneous sensor network3}
\end{figure}
\begin{figure}[!t]
\centering
\includegraphics[width=4.5in]{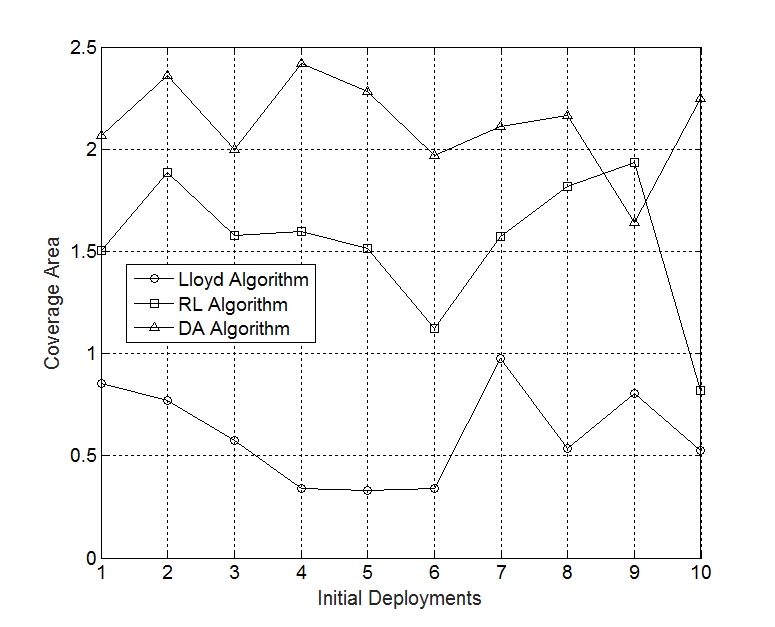}
\setlength{\belowcaptionskip}{-10pt} \caption{Comparison of coverage area with different algorithms in WSN3.}
\label{coverage in heterogeneous sensor network3}
\end{figure}

Figs. \ref{distortion in heterogeneous sensor network2} and \ref{coverage in heterogeneous sensor network2} illustrate the performance of the above algorithms for 10 random initial deployments in WSN2. Figs. \ref{distortion in heterogeneous sensor network3} and \ref{coverage in heterogeneous sensor network3} show similar performances in WSN3. The trends in  heterogeneous WSNs 2 and 3 are similar to those in homogeneous WSN1.

\section{Conclusions}
We studied the deployment of sensors in heterogeneous wireless sensor networks.
Similar to homogeneous WSNs, the
necessary condition for optimal deployment implies that every sensor node location
should coincide with the centroid of its own optimal sensing
region. Moreover, we considered a limited
communication range for the sensor nodes and modeled the sensor deployment
problem as a source coding problem with distortion reflecting
sensing accuracy. By defining an appropriate distortion measure, we
proposed a Restrained Lloyd algorithm and a Deterministic
Annealing algorithm to optimize sensor deployment in both
homogeneous and heterogeneous WSNs. Our simulation results
show that both DA and RL algorithms outperform the existing
Lloyd algorithm when communication range is limited and provide a fully connected network. The DA is not sensitive to initial conditions.

\appendices
\section{Proof of Proposition 1}
\begin{IEEEproof}
Let $\mbox{Part}_{ij}=\{q|\ \eta_i\|q-p_i\|^2\le\eta_j\|q-p_j\|^2\}$ be the pairwise weighted Voronoi region of Sensor $i$ when we only consider Sensors $i$ and $j$. Then, the exact weighted Voronoi region of Sensor $i$ is the intersection of these pairwise weighted Voronoi regions, i.e., $V^H_i(P)=\bigcap_{j\ne i}{\mbox{Part}_{ij}}$.
We define the coordinates of $q=(x,y)$, $p_i=(p_{ix},p_{iy})$ and $p_j=(p_{jx},p_{jy})$ and define $\eta=\sqrt{\eta_i/\eta_j}>0$. Then, expanding the hyperplane equation
$\eta_i\|q-p_i\|^2 = \eta_j\|q-p_j\|^2$
results in
\begin{equation}
\begin{aligned}
(\eta^2-1)(x^2+y^2)+2(p_{jx}-\eta^2p_{ix})x+2(p_{jy}-\eta^2p_{iy})y = p_{jx}^2-\eta^2p_{ix}^2+p_{jy}^2-\eta^2p_{iy}^2\\
\end{aligned}
\end{equation}
When $\eta=1$, the hyperplane equation is
\begin{equation}
2(p_{jx}-p_{ix})x+2(p_{jy}-p_{iy})y+p_{ix}^2+p_{iy}^2-p_{jx}^2-p_{jy}^2=0
\end{equation}
This hyperplane is the boundary of the half space $HS(p_i,p_j)$.
When $\eta<1$,  $\mbox{Part}_{ij}$ is defined by
\begin{equation}
(q-c)^2\geq r^2.
\end{equation}
When $\eta>1$,  $\mbox{Part}_{ij}$ is defined by
\begin{equation}
(q-c)^2\leq r^2,
\end{equation}
where $c=(\frac{p_{jx}-\eta^2p_{ix}}{1-\eta^2},\frac{p_{jy}-\eta^2p_{iy}}{1-\eta^2})$, $r=\frac{\eta}{|1-\eta^2|}\|p_i-p_j\|$.
Therefore, we have
\begin{equation}
\begin{aligned}
V^H_k(P) {=}&\:\{q \in Q | \eta_k\|q-p_k\| \le \eta_t\|q-p_t\|^2, \forall t \in 1,\cdots,n\}\\
         {=}&\:\left[\bigcap_{i:\eta_i<\eta_k}{B(c_{ik},r_{ik})}\right]\bigcap\left[\bigcap_{l:\eta_l=\eta_k}{HS(p_k,p_l)}\right]\bigcap\left[\bigcap_{j:\eta_j>\eta_k}{\left[B(c_{kj},r_{kj})\right]^c}\right]\\
         {=}&\:\left[\bigcap_{i:\eta_i<\eta_k}{B(c_{ik},r_{ik})}\right]\bigcap\left[\bigcap_{l:\eta_l=\eta_k}{HS(p_k,p_l)}\right]\bigcap\left[\bigcup_{j:\eta_j>\eta_k}{B(c_{kj},r_{kj})}\right]^c,\\
         \end{aligned}
\end{equation}
where $A^c$ denotes the complementary set of $A$.
\end{IEEEproof}

\section{Proof of Proposition 2}
\begin{IEEEproof}
Let $\bar{V}^H_k(P)$ be the weighted Voronoi region of Sensor $k$ when we ignore sensors with larger cost parameters. Accordingly, $\bar{V}^H_k(P)$ is defined by
\begin{equation}
\begin{aligned}
\bar{V}^H_k(P)=&\:\{q\in Q|\ \eta_k\|q-p_k\|^2\le\eta_i\|q-p_i\|^2,\text{for any} \ i\  \text{such that}\ \eta_k\ge\eta_i\}\\
\end{aligned}
\end{equation}
Review the definition of $V^H_k(P)$ in Eq. (\ref{df weighted Voronoi partition}), we have
\begin{equation}
\begin{aligned}
V^H_k(P)=&\:\{q\in Q|\ \eta_k\|q-p_k\|^2\le\eta_i\|q-p_i\|^2,
\eta_k\|q-p_k\|^2\le\eta_j\|q-p_j\|^2,\\
&\text{for any} \ i,j\  \text{such that}\ \eta_j>\eta_k, \eta_k\ge\eta_i\}\\
\end{aligned}
\end{equation}
Obviously, $\bar{V}_k(P)$ is the intersection of convex regions and therefore star-shaped. The relationship between $\bar{V}^H_i(P)$ and $V^H_i(P)$ is
\begin{equation}
\label{16}
\bar{V}^H_k(P) = V^H_k(P)\bigcup W_k,
\end{equation}
where $W_k = \{q\in Q|\ \eta_k\|q-p_k\|^2\ge\eta_j\|q-p_j\|^2,\eta_k\|q-p_k\|^2\le\eta_i\|q-p_i\|^2\,\text{for any} \ i,j\  \text{such that}\ \eta_j>\eta_k, \eta_k\ge\eta_i\}$.
For any point $q$ such that $|\eta_j\|q-p_j\|^2\le\eta_k\|q-p_k\|^2$ and $\eta_k\|q-p_k\|^2\le\eta_i\|q-p_i\|^2$, we have $\eta_j\|q-p_j\|^2\le\eta_i\|q-p_i\|^2$. So $W_k$ can be rewritten as
\begin{equation}
\label{17}
\begin{aligned}
W_k = &\:\{q\in Q|\ \eta_j\|q-p_j\|^2\le\eta_i\|q-p_i\|^2,\eta_j\|q-p_j\|^2\le\eta_k\|q-p_k\|^2,
\eta_k\|q-p_k\|^2\le\eta_i\|q-p_i\|^2\,\\
&\:\text{for any} \ i,j\  \text{such that}\ \eta_j>\eta_k, \eta_k\ge\eta_i\}.\\
\end{aligned}
\end{equation}
Due to the definitions of $\bar{V}^H_k(P)$ and $V^H_k(P)$, we have
\begin{equation}
\label{18}
\begin{aligned}
W_k&{=}\:\left[\bigcup_{j:\eta_j>\eta_k}V^H_j(P)\right]\bigcap\bar{V}^H_k(P) = \bigcup_{j:\eta_j>\eta_k}\left[V^H_j(P)\bigcap\bar{V}^H_k(P)\right]\\
\end{aligned}
\end{equation}
Replacing $W_k$ from Eq. (\ref{18}) in Eq. (\ref{16}), we get the final result
\begin{equation}
\label{20}
\bar{V}^H_k(P)=V^H_k(P)\bigcup_{j:\eta_j>\eta_k}\left[V^H_j(P)\bigcap\bar{V}^H_k(P)\right]
\end{equation}
The elements in the right side are disjoint subsets.
Without loss of generality, let us assume that the $m$ disjoint cost parameters are ordered such that the $k$-level cost parameter is larger than the $k+1$-level cost parameter. We also call the set including the indices of all sensors with a $k$-level cost parameter $Z_k$. Then:  \\
(1) For any level-1 sensor i, $V^H_i(P)$ is a convex set. Accordingly, the intersection $V^H_i(P)\bigcap W$ is a convex set and therefore Eq. (\ref{EQ1}) holds due to Proposition A.1 in \cite{SD}.\\
(2) Assume the equation holds for any sensor whose level is smaller than or equal to $k$. Consider a sensor $i$ whose level is $k+1$, by using the relationship between $\bar{V}^H_i(P)$ and $V^H_i(P)$ in Eq. (\ref{20}), we rewrite the objective function as
\begin{equation}
\begin{aligned}
\label{27}
&\int_{V^H_i(P)\bigcap W}\varphi(q,p_i)dq = \int_{\bar{V}^H_i(P)\bigcap W}\varphi(q,p_i)dq - \sum_{l=1}^{k}\sum_{t\in Z_l}\left[\int_{V^H_t(P)\bigcap\bar{V}^H_i(P)\bigcap W}\varphi(q,p_i)dq\right].
\end{aligned}
\end{equation}
Since $V^H_i(P)\bigcap W$ is a convex set and therefore star-shaped, the partial derivative of the first term can be solved by proposition A.1 in \cite{SD}. Sensors' levels in the second term are smaller than $k$ and thus the partial derivative of the second term can be solved by our assumption in Step (2). Therefore, the partial derivative becomes
\begin{equation}
\label{EQ28}
\begin{aligned}
&\frac{\partial \int_{V^H_i(P)\bigcap W}\varphi_i(q,p_i)dq}{\partial p_j}\\
=& \int_{\bar{V}^H_i(P)\bigcap W}\frac{\partial\varphi_i(q,p_i)}{\partial p_j}dq + \int_{\partial\left[\bar{V}^H_i(P)\bigcap W\right]}\varphi_i(\gamma,p_i)n^t(\gamma)\frac{\partial\gamma}{\partial p_j}d\gamma\\
&{-}\:\sum_{l=1}^{k}\sum_{t\in Z_l}\left[\int_{V^H_t(P)\bigcap\bar{V}^H_i(P)\bigcap W}\frac{\partial\varphi_i(q,p_i)}{\partial p_j}dq\right]\\
&{-}\:\sum_{l=1}^{k}\sum_{t\in Z_l}\left[\int_{\partial[V^H_t(P)\bigcap\bar{V}^H_i(P)\bigcap W]}\varphi_i(\gamma,p_i)n^t(\gamma)\frac{\partial\gamma}{\partial p_j}d\gamma\right].
\end{aligned}
\end{equation}
Note that
$\bar{V}^H_i(P)\bigcap W = \left[V^H_i(P)\bigcap W\right]\bigcup G(P)$,
where $G(P)=\left[\bigcup_{l=1}^{k}\bigcup_{t\in Z_l}(V_t(P)\bigcap V^H_i(P)\bigcap W)\right]$,
is a star-shaped set consisting of several disjoint subsets.
By using Lemma \ref{Lamma2}, we have
\begin{equation}
\begin{aligned}
&\int_{\partial\left[\bar{V}^H_i(P)\bigcap W\right]}\varphi_i(\gamma,p_i)n^t(\gamma)\frac{\partial\gamma}{\partial p_j}d\gamma - \sum_{l=1}^{k}\sum_{t\in Z_l}\left[\int_{\partial[V^H_t(P)\bigcap\bar{V}^H_i(P)\bigcap W]}\varphi_i(\gamma,p_i)n^t(\gamma)\frac{\partial\gamma}{\partial p_j}d\gamma\right]\\
=& \int_{\partial\left[V^H_i(P)\bigcap W\right]}\varphi_i(\gamma,p_i)n^t(\gamma)\frac{\partial\gamma}{\partial p_j}d\gamma.
\end{aligned}
\label{part1_19}
\end{equation}
Also, we have
\begin{equation}
\begin{aligned}
& \int_{\bar{V}^H_i(P)\bigcap W}\frac{\partial\varphi_i(q,p_i)}{\partial p_j}dq - \sum_{l=1}^{k}\sum_{t\in Z_l}\left[\int_{V^H_t(P)\bigcap\bar{V}^H_i(P)\bigcap W}\frac{\partial\varphi_i(q,p_i)}{\partial p_j}dq\right] = \int_{V^H_i(P)\bigcap W}\frac{\partial\varphi_i(q,p_i)}{\partial p_j}dq.
\end{aligned}
\label{part2_19}
\end{equation}
Eq. (\ref{EQ1}) is derived by replacing Eqs. (\ref{part1_19}) and (\ref{part2_19}) in Eq. (\ref{EQ28}).
In other words, Eq. (\ref{EQ1}) is correct for sensors whose level is smaller than or equal to $k+1$.
In summary, Eq. (\ref{EQ1}) is correct for sensors in all levels of heterogeneous WSNs.
\end{IEEEproof}

%




\section*{Acknowledgment}

The authors would like to thank Dr. Erdem Koyuncu for helpful discussions.

\ifCLASSOPTIONcaptionsoff
  \newpage
\fi

\end{document}